\documentclass[preprint,aps,prl,showpacs,twocolumn,mamsmath,10pt]{revtex4}
\usepackage{epsfig,epsf,graphics,psfrag}
\usepackage{times}
\usepackage{float}
\usepackage{color}
\usepackage{amsfonts,amssymb,stmaryrd,latexsym,amsmath}
\usepackage{hhline}

\newcommand{\be}{\begin{equation}}
\newcommand{\ee}{\end{equation}}
\newcommand{\ba}{\begin{eqnarray}}
\newcommand{\ea}{\end{eqnarray}}

\begin{document}
\preprint{\small FZJ-IKP-TH-2009-13, HISKP-TH-09/16}
\title{Implications of heavy quark spin symmetry on heavy meson hadronic
molecules}
\author{Feng-Kun Guo$^1$\footnote{{\it E-mail address:}
f.k.guo@fz-juelich.de}, Christoph~Hanhart$^{1,2}$\footnote{{\it
E-mail address:} c.hanhart@fz-juelich.de}, and Ulf-G.
Mei{\ss}ner$^{1,2,3}$\footnote{{\it E-mail address:}
meissner@itkp.uni-bonn.de}}

\affiliation{$\rm ^1$Institut f\"{u}r Kernphysik and J\"ulich Center
             for Hadron Physics, Forschungszentrum J\"{u}lich,
             D--52425 J\"{u}lich, Germany}%
\affiliation{$\rm ^2$Institute for Advanced Simulations,
             Forschungszentrum J\"{u}lich, D--52425 J\"{u}lich, Germany}
\affiliation{$\rm ^3$Helmholtz-Institut f\"ur Strahlen- und
             Kernphysik and Bethe Center for Theoretical Physics,\\ Universit\"at
             Bonn,  D--53115 Bonn, Germany}

\begin{abstract}
\noindent
  In recent years, many heavy mesons and charmonia were observed which
  do not fit in the conventional quark model expectations. Some of
  them are proposed to be hadronic molecules. Here we investigate the
  consequences of heavy quark spin symmetry on these heavy meson
  hadronic molecules. Heavy quark spin symmetry enables us to predict
  new heavy meson molecules and provides us with a method to test heavy
  meson molecule assumptions of some newly observed states.
  In particular, we predict an $\eta_c'f_0(980)$
  bound state as the spin-doublet partner of the $Y(4660)$ proposed as
  a $\psi'f_0(980)$ bound state with a mass of $4616^{+5}_{-6}$~MeV
  and the prominent decay mode $\eta_c'\pi\pi$. The width is predicted
  to be $\Gamma(\eta_c'\pi\pi)=60\pm30$~MeV. The $\pi^+\pi^-$
  invariant mass spectrum and the line shape  are
  calculated. We suggest to search for this state in $B^{\pm}\to
  \eta_c'K^{\pm}\pi^+\pi^-$, whose branching fraction is expected to be
  large.
\end{abstract}

\maketitle

\vspace{1cm}

The chromo-magnetic interaction of a heavy quark with gluons is
proportional to the magnetic moment of the heavy quark, and is
suppressed by the heavy quark mass (for reviews of heavy quark
symmetry, see Refs.~\cite{Neubert:1993mb,manoharbook}). Therefore in
the heavy quark limit $m_Q\to\infty$, the interaction is
spin-independent, and a new symmetry appears called heavy quark spin
symmetry. Due to this symmetry, there are spin multiplets of both
heavy mesons and heavy quarkonia, as e.g. the $\{D,D^*\}$ and
$\{\eta_c,J/\psi\}$. The masses of the members within the same spin
multiplet would be degenerate in the heavy quark limit. In this
Letter, we extend this symmetry to possible heavy meson molecules,
which were observed in recent years (for reviews, see e.g.
Ref.~\cite{charmreview1,charmreview2,charmreview3,charmreview4}).
By heavy meson molecules, we mean bound
states consisting a heavy meson/heavy quarkonium and a light
hadron, or two heavy mesons. In this Letter, we will focus on the
former type.

As an example, let us focus on the $D_{s0}^*(2317)$ and the
$D_{s1}(2460)$ first. They were proposed to be S-wave hadronic
molecules whose components are mainly $DK$ and $D^*K$,
respectively~\cite{ds1,ds2,ds3,ds4,ds5,ds6} (for the latest development on the
$D_{s0}^*(2317)$ in the hadronic molecular picture, we refer to
Refs.~\cite{ds23171,ds23172,ds23173,ds23174}). Their masses are measured to
be~\cite{Amsler:2008zz}
\ba%
M_{D_{s0}^*(2317)} &\!\!=&\!\! 2317.8\pm0.6~{\rm MeV}, \nonumber\\
M_{D_{s1}(2460)} &\!\!=&\!\! 2459.6\pm0.6~{\rm MeV}.
\ea%
Were they the bound states of $DK$ and $D^*K$, respectively, the
binding energies are \ba%
\epsilon_{D_{s0}^*(2317)} &\!\!=&\!\! M_D+M_K-M_{D_{s0}^*(2317)}=
45~{\rm MeV}, \nonumber\\
\epsilon_{D_{s1}(2460)} &\!\!=&\!\! M_{D^*}+M_K-M_{D_{s1}(2460)} =
45~{\rm MeV}, \ea%
where we have taken the averaged masses within the same isospin
multiplets of $D,D^*$ and $K$. One notices that the binding energies
are the same. For molecular states this appears to be natural: first
of all the leading interactions of light mesons with $D$ and $D^*$
mesons are independent of the heavy quark spin and secondly the
light meson--$D^{(*)}$ meson Greens functions, which provide an
important input to the bound-state equations, are to a very good
approximation mass-independent as long as evaluated close to the
corresponding threshold. As a result, hadronic molecules also fall
in spin multiplets, and, most importantly here, the splitting within
one multiplet remains the same as the hyperfine splitting between
the heavy mesons which are the components of the hadronic molecule.

In the same way the hyperfine splitting within a heavy quarkonium spin
multiplet will also be untouched by the interactions with light
mesons. The interaction between a heavy quarkonium and light hadrons
occur mainly through exchanging soft gluons. The leading order heavy
quarkonium interaction with a soft gluon field comes from the
chromo-electric dipole interaction~\cite{qcdme1,qcdme2,qcdme3} which is
spin-independent. The chromo-magnetic interaction is suppressed by
$1/m_Q$~\cite{Voloshin:2006ce}. Since any hadron should be color
singlet, the number of the exchanged soft gluons should be at least
two. Therefore, the suppression of the spin-dependent interactions
between a heavy quarkonium and light hadrons is at least $1/m_Q^2$.
As a result, a bound-state of a heavy quarkonium and light hadrons,
called hadro-charmonium in Ref.~\cite{Dubynskiy:2008mq}, will have
partner(s) whose components are the same light hadrons and the
spin-multiplet partner(s) of the same heavy quarkonium. The mass
splitting within the molecular spin-multiplet will be, to a very good
approximation, the same as the heavy quarkonium hyperfine splitting.

This nice feature enables us to predict new heavy meson molecules
and provides us with a method to test heavy meson molecule
assumptions of some newly observed states as illustrated in the
following.

The Belle Collaboration~\cite{Wang:2007ea} observed a resonant
structure, called $Y(4660)$, in the $\psi'\pi^+\pi^-$ final state
using the method of initial state radiation. The line shape of the
state was fitted with a P--wave Breit-Wigner~\footnote{An additional
  momentum factor had to be introduced in this analysis to account for
  the apparent asymmetry in the spectral function. This asymmetry
  emerges naturally in the analysis of Ref.~\cite{Guo:2008zg}.} in
the experimental paper, and they got $4664\pm12$~MeV. In
Ref.~\cite{Guo:2008zg}, we demonstrated that the experimental data
support a $\psi'f_0(980)$ bound state hypothesis for the $Y(4660)$.
As a result of fitting the mass, which gave
$M_Y=4665^{+3}_{-5}$~MeV, we could calculate the spectral
distribution. If this  interpretation of the $Y(4660)$ is indeed
correct, heavy quark spin symmetry implies that there is an
$\eta_c'f_0(980)$ bound state, to be called $Y_\eta$ in the
following. The quantum numbers of such a state are $J^P=0^-$.  The
$\eta_c'$ and $\psi'$ lie in the same spin multiplet, and their mass
splitting is
\be%
\Delta M = M_{\psi'}-M_{\eta_c'} = 49\pm4~{\rm MeV}.  \ee%
From the above analysis, the mass of the $\eta_c'f_0(980)$ bound state
would be \be%
\label{eq:my} M_{Y_\eta} = M_{Y(4660)} - \Delta M =
4616^{+5}_{-6}~{\rm MeV} \ , \ee%
where the uncertainties of the mass of the $Y(4660)$ and $\Delta M$
were added in quadrature.  Similar to the $Y(4660)$ decaying
predominantly into $\psi'\pi\pi$, the dominant decay channel of the
$Y_\eta$ would be $Y_\eta\to\eta_c'\pi\pi$.

If two particles form an S--wave bound state which is very close to
the threshold, there is a way to model-independently connect the
effective coupling constant of the bound state to its constituents,
$g$, directly to the molecular admixture of the
state~\cite{molecule1,molecule2}. Historically, Weinberg used this method to
show that the deuteron is not an elementary particle. Especially, one
may write for a pure molecule
\begin{eqnarray}
\frac{g^2}{4\pi}= 4(m_1+m_2)^2\sqrt{{2\epsilon\over\mu}} \left(1 +
{\cal O}\left({\sqrt{2\mu\epsilon}\over\beta}\right)\right)\ ,
\label{eq:geff}
\end{eqnarray}
where $m_1$ and $m_2$ denote the masses of the constituents,
$\epsilon$ the binding energy related to $M$, the mass of the molecule, via
$M=m_1+m_2-\epsilon$, $\mu=m_1m_2/(m_1+m_2)$ the reduced mass, and $1/\beta$
the range of the forces.

With the effective coupling constant fixed by Eq.~(\ref{eq:geff}),
we can predict the $\pi\pi$ invariant mass spectrum and the decay
width of the $Y_\eta\to\eta_c'\pi^+\pi^-$. Denoting the $\pi^+\pi^-$
invariant mass by $m_{\pi\pi}$, we have for the
differential width
\be%
\frac{d\Gamma_{Y_\eta}}{dm_{\pi\pi}^2} = \frac{g^2q}{8\pi M_{Y_\eta}^2}
\rho_{f_0}^{[\pi^+\pi^-]}(m_{\pi\pi}),
\ee%
where $q$ is the magnitude of the three-momentum of the $\eta_c'$ in
the $Y_\eta$ rest frame
$$q=\frac{\sqrt{\left[M_{Y_\eta}^2-(m_{\pi\pi}+M_{\eta_c'})^2
\right] \left[M_{Y_\eta}^2-(m_{\pi\pi}-M_{\eta_c'})^2
\right]}}{2M_{Y_\eta}},$$ and
$\rho_{f_0}^{[\pi^+\pi^-]}(m_{\pi\pi})$ is the $\pi^+\pi^-$ fraction
of the $f_0$ spectral function,
\be%
\rho_{f_0}^{[\pi^+\pi^-]}(m_{\pi\pi}) = \frac1{\pi}\frac{{\rm
Im}(\Pi^{\pi^+\pi^-}_{f_0}(m_{\pi\pi}))} {\left|
m_{\pi\pi}^2-m_{f_0}^2+\sum_{ab}\hat
\Pi^{ab}_{f_0}(m_{\pi\pi})\right|^2} \ ,
\ee%
where $\hat \Pi^{ab}_{f_0}(m_{\pi\pi})=\Pi^{ab}_{f_0}(m_{\pi\pi})
-{\rm
  Re}(\Pi^{ab}_{f_0}(m_{f_0}))$ denote the renormalized self-energies
of the $f_0$ with respect to the channel $ab=\pi\pi$ or $K{\bar K}$.
Analytic expressions are given in Ref.~\cite{achasov}. The imaginary
part of the self-energy of the $f_0$ is fixed by unitarity
\be%
{\rm Im}(\Pi^{\pi^+\pi^-}_{f_0}(m_{\pi\pi})) =
m_{f_0}\Gamma_{f_0\to\pi^+\pi^-}(m_{\pi\pi}),
\ee%
and
\be%
\Gamma_{f_0\to\pi^+\pi^-}(m_{\pi\pi}) = \frac{g_{f_0\pi^+\pi^-}^2}{16\pi
m_{f_0}}\sqrt{1-{4m_\pi^2\over m_{\pi\pi}^2}}.
\ee%
\begin{figure}[t]
\begin{center}
\vglue-10mm
\includegraphics[width=0.5\textwidth]{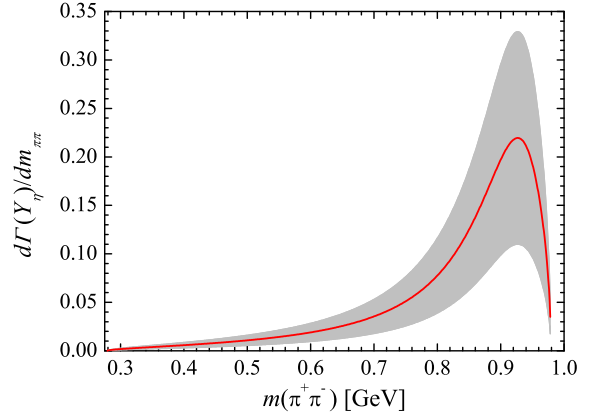}
\vglue-5mm \caption{The $\pi^+\pi^-$ invariant mass spectrum for the
$Y_\eta\to\eta_c'\pi^+\pi^-$ decay.\label{fig:pipi}}
\end{center}
\end{figure}

The input parameters related to the $f_0(980)$ are taken from the
fits provided in Ref.~\cite{kloe}.  To be specific, we use
$m_{f_0}=0.9862$~GeV, $g_{f_0K^+K^-}=3.87$~GeV, and
$g_{f_0\pi^+\pi^-}=-2.03$~GeV, which are the central values of the
various parameters of fit $K2$ shown in Table 4 of that reference.
The couplings for the neutral channels are fixed using the isospin
relations. With these parameters, and
$M_{\eta_c'}=3637\pm4$~MeV~\cite{Amsler:2008zz}, the $\pi^+\pi^-$
invariant mass spectrum for the $Y_\eta$ decaying into
$\eta_c'\pi^+\pi^-$ is predicted in Fig.~\ref{fig:pipi}, where the
solid line shows the result using the central values of
$M_{\eta_c'}$ and $M_{Y_\eta}$.

Integrating over the $\pi\pi$ invariant mass, and considering both
the $\eta_c'\pi^+\pi^-$ and $\eta_c'\pi^0\pi^0$ channels, we get a
width of the $Y_\eta\to\eta_c'\pi\pi$ as $58\pm5$~MeV. Here we only
took the leading term of the effective coupling constant as given in
Eq.~(\ref{eq:geff}), and the uncertainty comes solely from that of
the masses of the $Y_\eta$ and $\eta_c'$. However, to get a more
realistic estimate of the uncertainty, we need to also estimate,
e.g., higher order terms of the effective coupling constant. As
stated before, the interactions between a charmonium and light
hadrons are mediated by soft gluons, so we may estimate the range of
forces as $1/\beta\approx1/\Lambda_{\rm QCD}$. Noticing
$\sqrt{2\mu\epsilon}\approx100$~MeV, the uncertainty of $g^2$ from
this source, and hence the width, is about 50\%.  We want to stress
that this estimate is clearly quite conservative.  Taking the
expected inverse mass of the lightest glueball as the range of
forces might appear equally justified (and this would lead to a much
smaller uncertainty).  On the other hand we do not
explicitly include uncertainties from other sources like higher
orders in the $1/m_Q$ expansion.  Thus, as result, we get
\be%
\Gamma(Y_\eta\to\eta_c'\pi\pi) = 60\pm30~{\rm MeV} \ ,
\ee%
and the uncertainty in the $\pi^+\pi^-$ invariant mass spectrum is
reflected as the band in Fig.~\ref{fig:pipi}. The uncertainty in the
coupling constant has a much larger effect on the signal than the one
of the mass of the $Y_\eta$.

Assuming the width of the $Y_\eta$ being saturated by the
$\eta_c'\pi\pi$ final state, the line shape of the $Y_\eta$ can also be
predicted. For this we use a dispersion integral, which gives us an
expression for the $Y$ self-energy, $\Pi_{Y_\eta}(M)$, for arbitrary
values of $M$
\begin{equation}
\Pi_{Y_\eta}(M) = \frac{1}{\pi}\int_{M_{\rm thr}^2}^\infty \!\!\!\!
ds\frac{M_{Y_\eta}\Gamma^{\rm tot}_{Y_\eta}(\sqrt{s})}{s-M^2-i\epsilon} \ ,
\end{equation}
where $M_{\rm thr}=M_{\eta_c'}+2m_\pi$ denotes the lowest physical
threshold of relevance here and $\Gamma^{\rm
tot}_{Y_\eta}(\sqrt{s})$ the total width of the $Y_\eta$ as a
function of $\sqrt{s}$. Note that this treatment is completely
consistent to what was done for the $f_0$. With the self-energies at
hand we may now give the expression for the spectral function of the
$Y_\eta(4616)$
\begin{equation}
\rho_{Y_\eta}(M)=\frac{M_{Y_\eta}\Gamma_{Y_\eta}^{\rm tot}(M)}{\left|M^2-M_{Y_\eta}^2+\hat \Pi_{Y_\eta}(M)
\right|^2} \
, \label{eq:ysf}
\end{equation}
where, as above, we defined $\hat \Pi_{Y_\eta}(M) = \Pi_{Y_\eta}(M)-{\rm
Re}(\Pi_{Y_\eta}(M_{Y_\eta}))$.
\begin{figure}[t]
\begin{center}
\vglue-10mm
\includegraphics[width=0.5\textwidth]{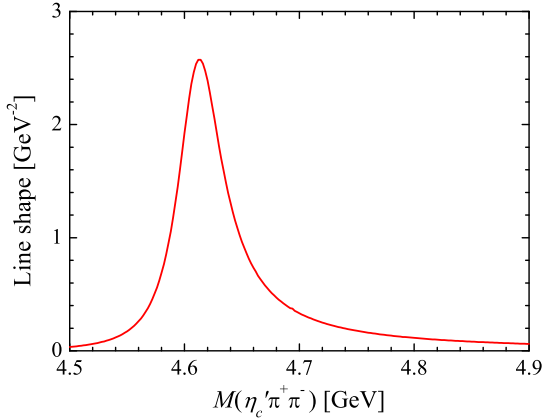}
\vglue-5mm \caption{Line shape of the $Y_\eta$ in the
$\eta_c'\pi^+\pi^-$ invariant mass
distribution.
For clarity we only show the line corresponding to the
central values of the parameter space. Note that the
asymmetry  seen in the distribution is an unavoidable
consequence of the molecular structure.
\label{fig:lineshape}}
\end{center}
\end{figure}
Replacing the total width of the $Y_\eta$ in the numerator of
Eq.~(\ref{eq:ysf}) by $\Gamma(Y_\eta\to\eta_c'\pi^+\pi^-)$, one gets
$\rho_{Y_\eta}^{[\eta_c'\pi^+\pi^-]}(M)$. That is the line shape of the
$Y_\eta$ in the $\eta_c'\pi^+\pi^-$ mass distribution as given in
Fig.~\ref{fig:lineshape}.

The proposed $\eta_c'f_0(980)$ bound state can be
searched for in $B$ decays. We suggest to search it in $B^{\pm}\to
\eta_c'K^{\pm}\pi^+\pi^-$. Taking data from
Ref.~\cite{Amsler:2008zz} for three measured channels, the branching
fraction of the so far unmeasured decay $B^{\pm}\to
\eta_c'K^{\pm}\pi^+\pi^-$ can be estimated as
\ba%
&& {\cal B}\left(B^{\pm}\to \eta_c'K^{\pm}\pi^+\pi^-\right) \nonumber\\
&&= {\cal B}\left(B^{\pm}\to \eta_c'K^{\pm}\right) \frac{{\cal
B}\left(B^{\pm}\to \psi'K^{\pm}\pi^+\pi^-\right)}{{\cal
B}\left(B^{\pm}\to \psi'K^{\pm}\right)} \nonumber\\
&&= (3.4\pm1.8)\times10^{-4}
\frac{(1.9\pm1.2)\times10^{-3}}{(6.48\pm0.35)\times10^{-4}}
\nonumber\\ &&\sim 1\times 10^{-3}.
\ea%
Such a large branching fraction offers a great opportunity of finding the
$Y_\eta$ in the $B$ decays, although we cannot predict ${\cal
B}\left(B^{\pm}\to Y_\eta K^{\pm}\right)$.

In summary, the heavy quark spin symmetry, which is exact in the
heavy quark limit, is extended to the systems made of a heavy
meson/quarkonium and light hadrons. We argue that the hyperfine
splitting remains untouched in heavy meson molecules. Based on this
observation, there should be an $\eta_c' f_0(980)$ bound state with
a mass of $4616^{+5}_{-6}$~MeV, were the $Y(4660)$ a $\psi'f_0(980)$
bound state as suggested in Ref.~\cite{Guo:2008zg}. Such a bound
state would decay mainly into $\eta_c'\pi\pi$ with a width of
$60\pm30$~MeV. In addition, analogous to the $Y(4660)$, we also
predict decays into $\eta_c'K^+K^-$ and $\eta_c'\gamma\gamma$. There
is also the possibility of a decay into
 $\Lambda_c^+\Lambda_c^-$.
We predict the $\pi^+\pi^-$ invariant mass spectrum of the
$Y_\eta(4616)$ decay into $\eta_c'\pi^+\pi^-$, and the prediction is
parameter-free. We also predict the line shape of the state in the
$\eta_c'\pi^+\pi^-$ final state assuming its decays are saturated by
the $\eta_c'\pi\pi$.

The state can be searched for in the $B$-factories Belle and BaBar.
The branching fraction of the $B^{\pm}\to\eta_c'K^{\pm}\pi^+\pi^-$ is
estimated to be of order $1\times 10^{-3}$, and hence there is a great
opportunity to find the $\eta_c'f_0(980)$ bound state proposed here in
the $\eta_c'\pi^+\pi^-$ final state of this decay. The state can also
be studied with ${\overline {\rm P}}$ANDA at FAIR~\cite{panda} in the future.
Such a study would be helpful to understand better not only the $XYZ$
states observed in recent years but also the interaction between a
charmonium and light hadrons, which can provide useful information for
understanding the charm production in relativistic heavy ion
collisions.

This work is partially supported by the Helm\-holtz Association
through funds provided to the virtual institute ``Spin and strong
QCD'' (VH-VI-231) and by the DFG (SFB/TR 16, ``Subnuclear Structure
of Matter''). We also acknowledge the support of the European
Community-Research Infrastructure Integrating Activity ``Study of
Strongly Interacting Matter'' (acronym HadronPhysics2, Grant
Agreement n. 227431) under the Seventh Framework Programme of EU.

\end{document}